\documentclass[aps,prb,twocolumn,groupedaddress]{revtex4}
\usepackage{epsfig,graphicx}
\usepackage[american]{babel}

\begin{document}

\title{Pressure dependence of the thermoelectric power of single-walled carbon nanotubes}
\author{
N. Bari\v{s}i\'{c}$^{1}$, 
R.Ga\'{a}l$^{1}$, 
I. K\'{e}zsm\'{a}rki$^{2}$, 
G. Mih\'{a}ly$^{2}$, 
L. Forr\'{o}$^{1}$}
\affiliation{
$^{1}$Institute of Physics of Complex Matter, Ecole Polytechnique Fédérale de Lausanne, CH-1015, Switzerland\\
$^{2}$Department of Physics, Technical University of Budapest, H-1111 Budapest, Hungary;}

\date{\today}

\begin{abstract}

We have measured the thermoelectric power~($S$) of high purity single-walled carbon nanotube mats as a function of temperature at various hydrostatic pressures up to 
\mbox{2.0 GPa}. 
The thermoelectric power is positive, and it increases in a monotonic way with increasing temperature for all pressures. The low temperature 
$({\rm T} < 40\ {\rm K})$ 
linear thermoelectric power is pressure independent and is characteristic for metallic nanotubes. At higher temperatures it is enhanced and though $S({\rm T})$ is linear again above about 
\mbox{100 K} 
it has a nonzero intercept. This enhancement is strongly pressure dependent and is related to the change of the phonon population with hydrostatic pressure.
\end{abstract}
\maketitle

Soon after the discovery of multi-walled carbon nanotubes (MWNT)\cite{1} 
single--walled carbon nanotubes (SWNT) were synthesised with catalytic metal particles.\cite{2} SWNT span electronic properties from metals to variable gap semiconductors, depending on the chirality and the diameter of the rolled up graphene sheet.\cite{3} It was immediately realized that metallic tubes represent the ultimate one--dimensional conductor, and they generated considerable interest due to the possible realization of the Luttinger liquid~(LL) behaviour in nature. Indeed, conductivity measurements performed on individual nanotubes have shown the canonical power--law behaviour of conductance as function of temperature and voltage.\cite{4} 

Early thermoelectric power~(TEP) experiments performed on nanotube bundles lead to qualitatively similar results:\cite{5}$^{-}$\cite{8} 
a metallic character due to hole carriers, a knee--feature (change of slope) around 
\mbox{100 K}, 
and preparation dependent magnitude. The observed TEP was associated with intrinsic electronic behaviour, and the fact that the nanotube bundles contain both metallic and semiconducting ropes allowed a description of the unusual shape of the $S({\rm T})$ curves in a simple two-band model, using a sufficient number of fitting parameters. It was also realized that in the high temperature range the thermoelectric power is significantly larger then expected for bundles constituted predominantly of metallic ropes.\cite{5}

Since SWNTs may be contaminated by magnetic impurities coming from the catalysts (Fe, Ni, Co) needed for their growth, the high value of the TEP above 100~K questions its intrinsic origin. The influence of the magnetic impurities has clearly been experimentally demonstrated:\cite{9} in the presence of transition metal impurities the thermopower has a strong peak, just around 100~K. This peak was associated with the Kondo scattering of conduction electrons on the impurity spins. The enhancement of the TEP could be largely, albeit not completely, removed by iodine neutralization.\cite{9} 
A non--intrinsic, residual "Kondo--term" may arise from magnetic impurities trapped on the body of the nanotubes, even in case of nominally pure samples.

The change of sign of the thermoelectric power upon degassing also raised the possibility of a non-intrinsic process.\cite{10} 
The conductivity of doped semiconducting tubes can be as high as that of the metallic ones and the overall thermopower may be determined by their contribution. Recently it was suggested that hole--doping due to oxygen adsorption explains the sign and the shape of $S({\rm T})$, at least in case of bundles where semiconducting ropes dominate.\cite{10}  

In this paper we report the results of thermoelectric power experiments on high purity 
single--wall nanotubes under hydrostatic pressures. We show that the low--temperature linear thermoelectric power arises from metallic ropes and it is consistent with the Luttinger liquid behaviour. Our results on pressurized samples exclude the "residual Kondo" origin of the high temperature TEP, and allow to separate the band and phonon contributions.  We propose a model, in which the knee--feature is coming from the change of the population of phonons that are relevant for the electron scattering. This picture is further supported by comparison of the thermoelectric power in carbon--based metallic compounds, which have a similar local bond structure but a quite different band structure.  

SWNT soot was purchased from Rice University, rich in catalytic particles. Most of these metallic blobs, covered with a graphitic shell, were eliminated using a purification method elaborated in our laboratory.\cite{11} 
This purification method results in open and less bundled, more dispersed nanotubes. The purified SWNTs from the suspension were deposited on a filtration membrane and a thin film of buckypaper was produced. The buckypaper was heated in vacuum to 
$1200\ ^{\circ}{\rm C}$ 
in order to eliminate absorbates. A thin slab of 
$3.0 \times 1.0 \times 0.01\ {\rm mm}^{3} $
was cut and mounted on a thermopower sample holder, which fits into a clamped pressure cell. Small metallic heaters installed at both ends of the sample generated the temperature gradient, which was measured with a Chromel--Constantan differential thermocouple. The pressure medium used in this study is kerosene and the maximum pressure is 2.0~GPa. The pressure was measured using a calibrated InSb pressure gauge. A similar sample arrangement and pressure cell was used in the study of the pressure dependence of the resistivity of SWNTs by Gaal et al.\cite{12}

\begin{figure}[tp]
\newlength{\sir} \newlength{\vis}
\setlength{\sir}{85mm}   
\setlength{\vis}{60.1mm} 
\centering\includegraphics[width=\sir,height=\vis]{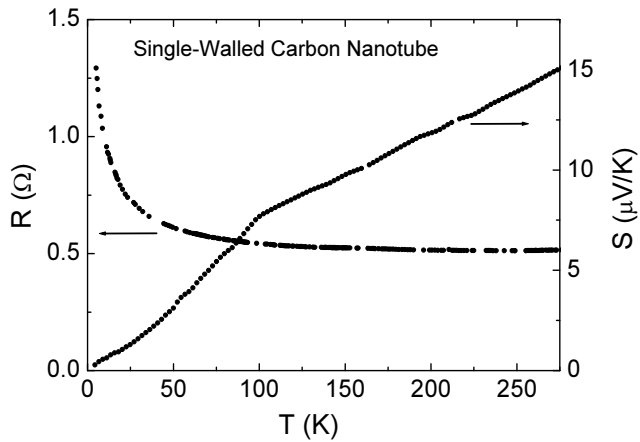}


\caption{The thermoelectric power $(S)$ and resistance versus temperature of a SWNT mat at ambient pressure. The SWNTs were purified from catalytic particles and heat treated at
$1200\ ^{\circ}{\rm C}$ in vacuum prior to measurement.}

\label{figure1}
\end{figure}

Figure~\ref{figure1} displays the resistance and the thermopower at ambient pressure in the 
${\rm T = 4 - 300\ K}$ 
region. The resistance has a negative slope in the whole temperature range, which is characteristic of an undoped sample. The low temperature increase is consistent with the signature of the LL feature.\cite{4} 
At low temperatures, below about 40~K, the thermopower is linear (see also the inset of Figure~{figure2}). The well--resolved linear TEP excludes the possibility of a gapped density of states, or hopping conduction trough mid--gap impurity levels, which might be invoked to account for the low temperature upturn in the resistance. At higher temperatures the thermoelectric power is enhanced, then its slope changes around 100~K. These features are in good qualitative agreement with earlier studies on pure nanotubes.\cite{5}$^{-}$\cite{8}   

\begin{figure}[tp]
\setlength{\sir}{85mm}   
\setlength{\vis}{60.1mm} 
\centering\includegraphics[width=\sir,height=\vis]{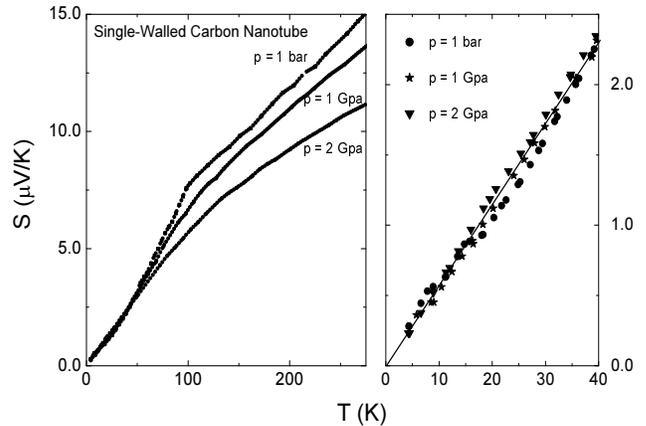}


\caption{The thermoelectric power of SWNT at 1bar, 10 and 20 kbar as a function of temperature. The low-temperature part is enlarged in the inset.}

\label{figure2}
\end{figure}

The temperature dependence of the TEP measured at various pressures is shown in Figure~\ref{figure2}. We found that the low--temperature linear part is not sensitive to the application of pressure, within the experimental resolution. In contrast at high temperatures the TEP is strongly pressure dependent. The distinct behaviour at low and high temperatures suggests that the thermoelectric power has different contributions in the two ranges. This observation excludes the possibility that the thermoelectric power is related to residual magnetic impurities. Pressure does not influence the sample purity or the exchange coupling~(J) between the conduction electrons and the magnetic impurity, thus it could modify a possible contribution of Kondo origin solely by shifting the Kondo temperature, 
\begin{equation}
T_{K}\approx \frac{E_{F}}{k_{B}} \cdot
\exp \left( -\frac{1}{2|J| D(E_{F})}\right).
\label{eq1}
\end{equation}
Here $D(\varepsilon _{F})$ is the electronic density of states at the Fermi level, which might have a pressure dependence. Such a pressure dependence should, however, also be reflected in the metallic low temperature range, which is not observed. Moreover, $T_{K}$ should decrease with increasing pressure as $D(\varepsilon _{F})$ decreases. Instead, one rather observes a smearing of the shoulder due to the application of the pressure than a structure, which shifts with pressure. 

The pressure study also shows that characteristic $S({\rm T})$ shape shown in Figure~\ref{figure1} and published by various groups\cite{5}$^{-}$\cite{8} cannot arise from oxygen adsorption, either. In the doped semiconductor picture the TEP is interpreted in terms of the particular shape of electronic density of states. Any pressure induced modification of the density of states would rescale the $S({\rm T})$ curve in the whole temperature range, in contrast to the observed differences in the low and high temperature ranges. 

The linear temperature dependence at low temperatures is characteristic of a simple metal, obeying Fermi liquid theory. Recent theoretical studies\cite{13,14} of the thermal transport in the Luttinger liquid phase showed, that  -- though both the electrical and the thermal conduction are unconventional -- the low temperature thermopower still follows Mott's formula of the logarithmic derivative of the conductivity, $\sigma $ ,
\begin{equation}
S=-\frac{\pi ^{2}}{3}\frac{k_{B}^{2}}{e}
\left( \frac{\partial \ln \sigma(\varepsilon )}{\partial \varepsilon} 
\right)_{\varepsilon =\varepsilon _{F}}.
\label{eq2}
\end{equation}
This leads to $S = cT$, whit a non-universal coefficient, c, which is determined by the band shape and the scattering rate. For the Luttinger liquid of spatial extension~L the (\ref{eq2}) holds only for low temperatures, while an enhancement is expected for
\( k_{B}T > \hbar v_{F}/gL \)    
($v_{F}$ is the Fermi velocity, while $g$ is the correlation parameter of the LL).\cite{14}  We do not believe this is the source of the observed enhancement in our case, since for realistic values of defect free nanotube segments 
$({\rm L} < 100\ {\rm nm} ) $
the corresponding temperature is above the range covered by the experiments.

The attenuation of the shoulder in $S({\rm T})$ with the pressure puts forward another explanation for the overall temperature dependence.\cite{8,15} If inelastic scattering of electrons on phonons is of importance, then the thermoelectric power is expressed as
\begin{equation}
S=S_{0}(1+\lambda \lambda _{S}({\rm T})),
\label{eq3}
\end{equation}
where $S_{0}$ is given by (\ref{eq2}), $\lambda $ is the standard mass enhancement due to renormalization by phonons and  $\lambda_{S}({\rm T})$ is the temperature--dependent enhancement of the thermopower. The latter quantity contains the Eliashberg function, and if the phonon density of states has a peak then  $\lambda_{S}({\rm T})$ will have a shoulder at the corresponding temperatures. This term gives a non--zero intercept of $S({\rm T})$ when extrapolated from high temperatures. Furthermore, this term can fade away thus changing the slope of $S$ when the population of phonons, which strongly scatter the charge carriers, diminishes. 

\begin{figure}[t]
\setlength{\sir}{85mm}   
\setlength{\vis}{115mm} 
\centering\includegraphics[width=\sir,height=\vis]{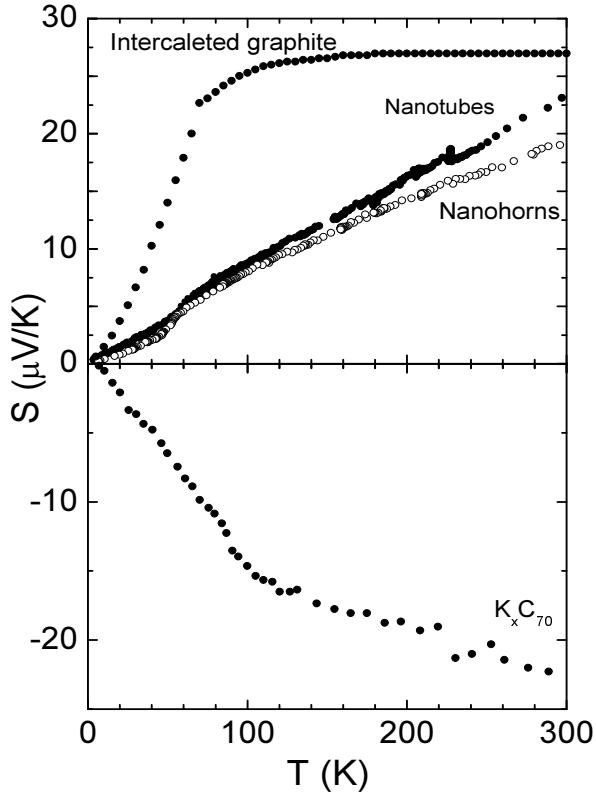}


\caption{The thermoelectric power of several carbon-based compounds, like intercalated graphite,\cite{16} carbon nanohorns,\cite{17} alkali metal doped C$_{70}$,\cite{18} and SWNTs. All the samples show similar break in slope in the 100~K range.}

\label{figure3}
\end{figure}

The model can also account for the observed pressure dependence. High pressure increases the force constant acting between the carbon atoms, shifting any feature in the phonon density of states to higher energies. Consequently, the population of the phonons, which strongly interact with the charge carriers, decreases and the thermopower drops. This explanation is corroborated by the fact that the temperature at which the shoulder in $S$ develops somewhat increases towards higher temperatures at high pressures.

The phonon mode in the case of SWNT should be the vibration of the carbon atoms in the hexagons, since similar feature in $S$ could be observed in other carbon--based materials, which are built up from the benzyl--ring units. Figure~\ref{figure3} summarizes the Seebeck coefficient of intercalated graphite,\cite{16} carbon nanohorns,\cite{17} alkali metal doped C$_{70}$,\cite{18} and SWNTs. All the samples show similar break in slope in the 100~K range. 

In conclusion, we have performed thermopower measurements on purified SWNT mat as a function of temperature and hydrostatic pressure. While the low temperature metallic thermopower was found to be pressure independent, the high temperature unusual TEP contribution was strongly suppressed by the application of pressure. We showed that the thermopower of our sample is due to metallic nanotubes, and we excluded non--intrinsic effects, like Kondo scattering of residual impurities or doping by adsorbed oxygen. Our results suggest that the temperature and pressure dependencies are governed by the change of population of phonons, connected with the thermal excitation of the hexagons (the building units of the SWNTs).

\begin{acknowledgments}
The sample purification by Klara Hernadi and useful discussions with Walter Schirmacher are acknowledged. This work has been supported by the Swiss National Foundation for Scientific Research and the NCCR Pool "MaNEP".
\end{acknowledgments}

\pagebreak

\end{document}